\documentclass[11pt, a4paper]{article}

\usepackage{jcappub}
\usepackage{amsmath, amsthm,amsfonts,rotate}
\usepackage{bm}
\usepackage{graphicx}
\usepackage{epstopdf}
\usepackage{hyperref}

\newcommand{\n}{{\bf \nabla}}

\newcommand{\LCDM}{\Lambda{\rm CDM}}

\newcommand{\tb}{{\bf \tau}}

\newcommand{\Op}{\mathcal{O}}
\newcommand{\PP}{{\bf \Pi}}
\newcommand{\VV}{{\bf V}}

\newcommand{\vv}{{\bf v}}
\newcommand{\rhO}{\rho_{\odot}}
\newcommand{\rO}{r_{\odot}}

\newcommand{\br}{{\bf r}}
\newcommand{\xx}{{\bf \xi}}

\newcommand{\zh}{\hat{z}}

\newcommand{\OO}{{\bf \Omega}}
\newcommand{\vr}{v_{\rm rot}}

\begin{document}

\title{Condensation of Galactic\\ Cold Dark Matter}

\author{Luca Visinelli}
\affiliation{Nordita, KTH Royal Institute of Technology and Stockholm University,\\ SE-106 91 Stockholm, Sweden}
\emailAdd{Luca.Visinelli@studio.unibo.it}

\abstract{
We consider the steady-state regime describing the density profile of a dark matter halo, if dark matter is treated as a Bose-Einstein condensate. We first solve the fluid equation for ``canonical'' cold dark matter, obtaining a class of density profiles which includes the Navarro-Frenk-White profile, and which diverge at the halo core. We then solve numerically the equation obtained when an additional ``quantum pressure'' term is included in the computation of the density profile. The solution to this latter case is finite at the halo core, possibly avoiding the ``cuspy halo problem'' present in some cold dark matter theories. Within the model proposed, we predict the mass of the cold dark matter particle to be of the order of $M_\chi c^2 \approx 10^{-24}$~eV, which is of the same order of magnitude as that predicted in ultra-light scalar cold dark matter models. Finally, we derive the differential equation describing perturbations in the density and the pressure of the dark matter fluid.}

\arxivnumber{}

\maketitle

\section{Introduction}

Observations on the rotational curves of spiral galaxies show that the velocities of the virialized material lying farther than the extent of the luminous matter from the galactic center reach a constant value~\cite{binney_book}. Various theories aim at explaining this discrepancy between observations and Newton's virial theorem, including a modification of the gravitational potential~\cite{sanders1984,sanders1986} or of the Poisson equation~\cite{milgrom1983, bekenstein1984}, conformal gravity~\cite{mannheim1993, mannheim1997}, and the metric skew tensor gravity~\cite{moffat1996, brownstein1, brownstein2}.

Nowadays, the most promising way to explain the observations of the galactic rotation curves~\cite{rubin1, rubin2} consists in postulating the existence of non-luminous (dark) matter, distributed in a halo which extends much farther than the luminous component of a galaxy. Further, this dark component is supposed to be non-relativistic (Cold Dark Matter, CDM~\cite{peebles1982, bond1982, blumenthal1982}, see also~\cite{peebles2003, padmanabhan2003}), since it is usually assumed to consist of massive particles with very low thermal velocities. Work on colliding galaxy clusters seem to confirm the existence of dark matter dominating the mass content of spiral galaxies and galaxy clusters~\cite{clowe, brada}. An indirect confirmation also comes from the success of the concordance cosmological model, or $\Lambda$ Cold Dark Matter ($\LCDM$) model, in reproducing the anisotropies observed in the cosmic microwave background~\cite{wmap, planck}. Among the most promising candidates for the CDM component are the Weakly Interacting Massive Particle (WIMP,~\cite{bertone2005}) or a population of zero-momentum axions~\cite{peccei, preskill1983, abbott1983, dine1983, stecker1983}.

However, these ``canonical'' CDM models usually feature problems in reproducing some observable properties of galaxies, most remarkably the under-abundance of small scale structure with respect to what predicted from simulations (the ``missing satellite'' problem~\cite{kauffmann1993, klypin1999, moore1999a}), the absence of a central density cusp rather predicted in numerical simulations (the ``cusp'' problem~\cite{navarro1997, moore1999b, ostriker2003, Romanowsky2003}), and too little dense subhalos compared with theoretical expectations (the ``too big to fail'' problem~\cite{kolchin2011}). In more detail, observations of both nearby dwarf galaxies and low surface brightness galaxies show that the density profile of the CDM halo at the core reaches a constant value~\cite{moore1994, burkert1995, persic1988, kravtsov1998, deblok2001, swaters2003, spekkens2005}. In contrast, various N-body simulations predict that the CDM density distribution steepens at the center of the halo~\cite{navarro1995, navarro1996, fukushige1997, ghigna1997, moore1998}.

Among the solutions proposed to overcome these issues, it has been suggested that dark matter could consist of a coherent scalar field with long range correlation, whose quanta are very light particles~\cite{sin, ji, sahni, hu2000, peebles2000}. In fact, on short length scales, light scalar fields do not behave as perfect CDM and would inhibit cosmological structure growth~\cite{marsh2013, hlozek2014, bozek2014}. This solution would consist of a viable mean to suppress low mass galaxies and provide cored profiles in CDM-dominated galaxies~\cite{hu2000, peebles2000, marsh2015}. This peculiar form of dark matter might form a Bose-Einstein Condensate (BEC), described by the Gross-Pitaevskii or non-linear Schr{\"o}dinger equation~\cite{goodman2000, riotto2000, chen2005, boehmer2007, brook2009, lee2009, kain2010, harko2011a, harko2011b, kain2012, harko2012, shapiro2012, guzman2013, harko2014, schroven2015}. Alternatively, axions can also be modeled as a coherent BEC with small spatial gradient~\cite{sikivie2009, park2012, pires2012, arias2012, banik2013, hlozek2015}. Recent 3D simulations of the gravitational collapse of wavelike cold dark matter ($\Psi$DM) with a mass of the order of $10^{-22}$eV show the effects on structure formation due to their large Jeans length~\cite{schive2014a, schive2014b, schive2016}, which improved over previous work on the subject~\cite{goodman2000, riotto2000, chen2005, boehmer2007}. Other alternative models embed dark matter condensation into space-time with torsion~\cite{fabbri2013}. A general review of the models proposed is given in Ref.~\cite{suarez2013}. The cosmological evolution of a BEC dark matter component has also been extensively explored~\cite{ferrer2004, grifols2006, fukuyama2008, fukuyama2009}.

At galactic scales, the evolution of a self-gravitating CDM system can be described as a fluid following the equation of continuity and the Navier-Stokes Equation (NSE). When these equations are derived for a Bose-Einstein fluid, an additional ``Quantum Pressure'' (QP) term appears~\cite{chen2005}.

In this paper, we treat cold dark matter as a pressure fluid with a dynamics described by the NSE, and we derive the equations for the zeroth- and first-order perturbations in the density and pressure of ``canonical'' and BEC cold dark matter. For this, we assume a rotating halo in which the proper velocity of dark matter is treated as a first order perturbation in the motion.

The paper is organized as follows. After the short review of fluid dynamics in Sec.~\ref{review}, we discuss some popular halo models fitting numerical simulations in Sec.~\ref{historicalprofiles}. In Sec.~\ref{canonicalCDM} we show that, at the lowest order, the dark matter density in the halo follows the Lane-Emden equation in a rotating frame. When quantum pressure is included, the Lane-Emden equation modifies as discussed in Sec.~\ref{BEC_CDM}, and results for the halo profile predict a finite core. In Appendix~\ref{pert_eq}, a generic expression for density perturbations and the proper velocity of dark matter is derived and will be part of future work.

\section{Equation for fluid dynamics} \label{review}
Newton's equations for a parcel of density $\rho$ and proper velocity $\vv$, written in a reference frame with the $\zh$ axis in the direction of increasing altitude, reads
\begin{equation} \label{newton}
\frac{d\vv}{dt} = -\frac{1}{\rho}\,\n\,p - \n\,\phi + \n\cdot \PP + \tb.
\end{equation}
Here, $p$ is the pressure acting on the parcel, $\phi$ is the gravitational potential, and $\tb$ describes all additional external forces in the system, like the mean gravitational field generated by all nearby galaxies. In addition, $\PP$ is a rank-two tensor describing the dissipative phenomena in the fluid,
\begin{equation} \label{newton_diss}
\n\cdot\PP = \eta\,\n^2\,\vv + \left(\zeta + \frac{\eta}{3}\right)\,\n\,\left(\n\cdot \vv \right).
\end{equation}
The two constants appearing in Eq.~\eqref{newton_diss} are known in the literature respectively as the dynamic viscosity $\eta$ and the second viscosity coefficient $\zeta$~\cite{beckett_book, bird_2007book}. The total time derivative of the velocity field can be explicitly written as the sum of a partial time derivative and the dyadic (advection) term which introduces a non-linear component in Newton's equation,
\begin{equation} \label{dyadic_equivalency}
\frac{d \vv}{dt} = \frac{\partial \vv}{\partial t} + \left(\vv \cdot \n\right) \,\vv = \frac{\partial \vv}{\partial t} + \n\,\left(\frac{v^2}{2}\right) - \vv\times\xx,
\end{equation}
where $v = |\vv|$ and we have defined the vorticity of the velocity field as
\begin{equation}
\xx = \n \times \vv.
\end{equation}
We assume that the galactic halo rotates at a constant rate $\OO$, and we switch to the rotating frame by setting $\vv \to \vv + \OO \times {\bf r}$, obtaining the NSE in the rotating frame
\begin{equation} \label{newton1}
\frac{\partial \vv}{\partial t} - \vv\times\xx = -\frac{1}{\rho}\,\n\,p - \n\,\left(\frac{v^2}{2} + \phi\right) -\OO \times \OO \times r - 2\,\OO\times \vv + \eta\,\n^2\,\vv.
\end{equation}
Here, $\OO \times \OO \times r$ and $2\,\OO\times \vv$ are respectively the Coriolis and the centrifugal acceleration terms. The NSE couples to two additional equations which express flux conservation (continuity equation),
\begin{equation} \label{continuity}
\frac{d\rho}{dt} +\rho\,\left(\n\cdot \vv\right) = 0,
\end{equation}
and the value of the gravitational potential generated by the matter density $\rho$ (Poisson equation),
\begin{equation} \label{poisson}
\n^2\phi = 4\pi\,G\,\rho.
\end{equation}
In the following, we look for a solution to the set of Eqs.~\eqref{newton1} and \eqref{continuity} in the steady-state regime,
\begin{equation} \label{del_velocity}
\frac{\partial \vv}{\partial t} = \frac{\partial \rho}{\partial t} = 0.
\end{equation}
Furthermore, since we are treating the velocity as a first order term in the perturbation series, we neglect all advection terms. Under these conditions, Eq.~\eqref{newton1} reads
\begin{equation} \label{newton1_steady}
\frac{1}{\rho}\,\n\,p + \n\,\left(\frac{v^2}{2} + \phi\right) + \OO \times \OO \times r + 2\,\OO\times \vv = \eta\,\n^2\,\vv.
\end{equation}
while Eq.~\eqref{continuity} in the steady-state regime is rewritten as the incompressibility relation $\n\cdot \vv = 0$ for the DM flow.

\section{Density profiles from N-body simulations} \label{historicalprofiles}

In the following, we define the profiles so that to match the local halo density $\rhO = 0.4{\rm ~GeV/cm^3}$~\cite{pato2015, xia2015, mckee2015, weber2009, catena_2009} (see also Ref.~\cite{bergstrom_1998}) when the radius $r$ equates the distance of the Sun from the center of the halo $\rO \approx 8.5$~kpc.

The $\LCDM$ model is very successful in predicting various observational features such as the galaxy rotation curves~\cite{rubin1, rubin2}, which are closely related to the nature of the CDM and the formation history of the halo. Observations show that the shape of rotation curves of CDM-dominated galaxies is universal across a wide mass range~\cite{persic1988, burkert1995, kravtsov1998}.

Early work on the analytical shape of the density profile~\cite{gunn1972, bahcall1980, fillmore1984, bertschinger1985, hoffman1985, hoffman1988}, showed that an isothermal profile with $\rho \sim r^{-2}$ is to be expected in a flat Friedmann-Robertson-Walker Universe. In addition, the density profile would steepen when sharper spectra of the initial density perturbation are assumed. These claims were first validated in simulations~\cite{frenk1985, quinn1986, zurek1988, warren1992, crone1994}.
The isothermal profile~\cite{bahcall1980} is derived from balancing the gravitational and pressure forces, assuming that pressure is linearly dependent on the matter density,
\begin{equation} \label{profile_isothermal}
\rho_{\rm iso}(r) = \rhO\,\frac{q_s^2+1}{q_s^2+q^2},
\end{equation}
where $q = r/\rO$, $q_s = r_s/\rO$, and $\rho_{\rm iso}$ and $r_s$ respectively parametrize density and radial size of the Galaxy. The isothermal profile has the characteristic to remain finite when $r\to 0$, approaching the value
\begin{equation} \label{profile_isothermal_core}
\rho_{\rm iso}(0) = \rhO\,\left(1+q_s^{-2}\right).
\end{equation}
However, the isothermal profile is too shallow when compared with results from numerical simulations of non-colliding dark matter. Departures from the power-law behavior were first reported in Refs.~\cite{frenk1988, efstathiou1988, dubinski1991}. A recent review has been given in Ref.~\cite{frenk2012}. Using numerical simulations, Navarro, Frenk, and White~\cite[NFW,][]{navarro1995, navarro1996, navarro1997} fitted results for CDM halos with mass spanning over four orders of magnitude with the function
\begin{equation} \label{profile_NFW}
\rho_{\rm NFW}(r) = \frac{\rhO}{q}\,\left(\frac{q_s+1}{q_s + q}\right)^2,
\end{equation}
thus predicting a rather dense central cusp growing with $\sim 1/q$.

Additional numerical simulations~\cite{ghigna1997, moore1998} motivated Moore~\cite{moore1999a, moore1999b} to propose the following form for the CDM profile
\begin{equation}
\rho_M(r) = \frac{\rhO}{q^{3/2}}\,\left(\frac{q_s + 1}{q_s + q}\right)^{3/2},
\end{equation}
which diverges at the core as $\sim 1/q^{3/2}$. The isothermal, NFW, and Moore profiles, together with other parametrizations such as the BE~\cite{binney2001} and PISO~\cite{deboer2005} profiles, can collectively be described by the generalized expression~\cite{diemand2004}
\begin{equation}
\rho_{\rm CDM}(q,q_s\alpha,\beta,\gamma) = \rhO\,\Theta_{\rm CDM}(q,q_s, \alpha,\beta,\gamma),
\end{equation}
with
\begin{equation} \label{profile_parametrized}
\Theta_{\rm CDM}(q, q_s, \alpha,\beta,\gamma) = q^{-\gamma}\,\left[\frac{q_s^\alpha+1}{q_s^\alpha+q^\alpha}\right]^{(\beta-\gamma)/\alpha}.
\end{equation}
The function $\Theta_{\rm CDM}(q, q_s, \alpha,\beta,\gamma)$ is normalized so that
\begin{equation}
\Theta_{\rm CDM}(1, q_s, \alpha,\beta,\gamma) = 1.
\end{equation}
In Table~\ref{table_profiles}, we have summarized the values of the parameters $\alpha$, $\beta$, and $\gamma$ for these popular halo models, including references.
\begin{table}[h!]
\centering
\begin{tabular}{|l|r|ccc|l|}
\hline
Profile & $r_s$(kpc) & $\alpha$ & $\beta$ & $\gamma$ & Reference\\
\hline
Isothermal & 5.0 & 2.0 & 2.0 & 0.0 & Bahcall~\cite{bahcall1980}\\
NFW & 20.0 & 1.0 & 3.0 & 1.0 & Navarro {\it et al.}~\cite{navarro1997}\\
Moore & 30.0 & 1.0 & 3.0 & 1.5 & Moore {\it et al.}~\cite{moore1999a, moore1999b}\\
BE & 10.2 & 1.0 & 3.0 & 0.3 & Binney and Evans~\cite{binney2001}\\
PISO & 4.0 & 2.0 & 2.0 & 0.0 & de Boer {\it et al.}~\cite{deboer2005}\\ 
\hline
\end{tabular}
\label{table_profiles}
\caption{Values of the parameters appearing in Eq.~\eqref{profile_parametrized} for various density profiles considered in the literature.}
\end{table}

More recent N-body simulations~\cite{navarro2004, merritt2006, gao2008, hayashi2008, springer_2008, stadel2009, navarro2010} favor three-parameter profile models like the Einasto profile~\cite{einasto1965, einasto1989}, a generalization over the S\'ersic model~\cite{sersic1963},
\begin{equation} \label{einasto_profile}
\rho_E(r) = \rhO\,\Theta_E(r),\quad\hbox{with}\quad \Theta_E(r) = {\rm exp}\left[-\frac{2}{\delta}\,q_s^{-\delta}\,\left(q^{\delta}-1\right)\right],
\end{equation}
where $r_s$ and $\delta$ are constants. The halo density at the core predicted by the Einasto profile is finite, with
\begin{equation} \label{einasto_profile_core}
\rho_E(0) = \rhO\,{\rm exp}\left(\frac{2}{\delta}\,q_s^{-\delta}\right).
\end{equation}

\begin{figure}[t!]
\centering
\includegraphics[height=10cm,width=10cm]{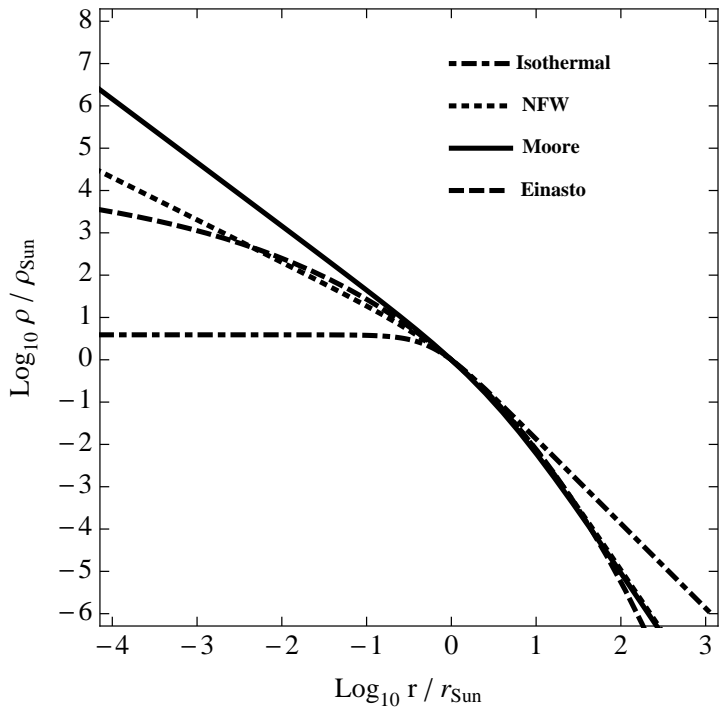}
\caption{The CDM profile from various fits to numerical simulations. Lines show the isothermal (dot-dashed line), NFW (dotted line), Moore (solid line), and Einasto profiles (dashed line).}
 \label{Fig_Literature}
\end{figure}

In Figure~\ref{Fig_Literature}, we compare the plots of the isothermal (dot-dashed line), NFW (dotted line), Moore (solid line), and Einasto profiles (dashed line), with the normalization satisfying $\Theta_{\rm CDM}(\rO) = 1$. For the profiles considered, we have set $r_s$ equal to the value given in Table~\ref{table_profiles}, while the parameters for the Einasto profile are $r_s = 20{\rm ~kpc}$ and $\delta = 0.17$~\footnote{A similar comparison has been presented in Refs.~\cite{lehoucq2009,weber2009}}.

\section{Equations for the halo profile neglecting quantum pressure corrections} \label{canonicalCDM}

We assume that the DM stream velocity is $v \ll \Omega\,L$, where $L$ is the typical galactic length scale. To give a numerical example, for a period of rotation $\Omega^{-1} = 200\,$My and for a length scale $L = 50\,$kpc, we obtain $v \ll 250$km/s. For this reason, we first look at a numerical resolution of the set of Eqs.~\eqref{newton1}-\eqref{poisson} in which we neglect the stream velocity. Under these conditions, the set of equations describing the balance between pressure and density in a galactic CDM halo is
\begin{eqnarray}
\n\,\phi &=& -\frac{1}{\rho}\,\n\,p - \OO \times \OO \times r, \label{NS_bulk1} \\
\frac{d\rho}{dt} &=& 0, \label{continuity1} \\
\n^2\phi &=& 4\pi\,G\,\rho \label{poisson1} .
\end{eqnarray}
Eq.~\eqref{continuity1} expresses mass conservation around an infinitesimal volume. The curl of Eq.~\eqref{NS_bulk1} yields $\n\rho \,\times \n\,p = 0$, which is a structural condition between the bulk pressure and density which is fulfilled by the barotropic relation
\begin{equation} \label{barotropic_relation}
p = p(\rho).
\end{equation}
Combining the divergence of the NS equation, first line in Eq.~\eqref{NS_bulk1}, and the Poisson equation, third line in Eq.~\eqref{NS_bulk1}, yields the equation
\begin{equation} \label{bulk_eq}
4\pi\,G\,\rho = -\n\left(\frac{1}{\rho}\,\n\,p\right) + 2\Omega^2,
\end{equation}
which is known in the literature as the Lane-Emden equation~\cite{lane, emden, chandrasekar}. For the generic barotropic relation expressed in Eq.~\eqref{barotropic_relation}, the Lane-Emden equation requires a numerical resolution~\cite{veugelen1985, kumar2011, alzate2014}. An analytic solution exists when the relation is polytropic $p \propto \rho^{1+1/n}$, with $n = 1, 2, 5$. When $\Omega = 0$, Eq.~\eqref{bulk_eq} has often found applications in the study of collision-less systems such as globular clusters and primordial galaxies~\cite{binney_book}. The rotating Lane-Emden equation in cylindrical coordinates has been discussed by Stodolkiewicz~\cite{stodolkiewicz1963} and Ostriker~\cite{ostriker1964} for the case of a non-rotating isothermal cylinder,  by Schneider and Schmitz~\cite{schneider} for a generic polytropic fluid, and by Christodoulou and Kazanas~\cite{christodoulou} in the context of planetary formation for a linear polytropic relation $p \propto \rho$ (see also Refs.~\cite{chavanis, martino}).

Here, we consider the case in which the rotation is not neglected, assuming that both density and pressure of DM in the galactic disk do not depend on the azimuthal coordinate $\phi$. We assume a cylindrical symmetry of the density, by setting
\begin{equation}
\rho = \rhO \,\Theta(r),
\end{equation}
where $\Theta = \Theta(r)$ is a function depending on the distance from the galactic center $r$ only, normalized so that $\Theta(\rO) = 1$. In cylindrical coordinates, Eq.~\eqref{bulk_eq} is expressed as
\begin{equation} \label{lane_emden}
4\pi\,G\,\rho + \frac{1}{r}\,\frac{d}{dr}\,\left(\frac{r}{\rho}\,\frac{dp}{dr}\right) = 2\Omega^2.
\end{equation}
To enforce the barotropic relation in Eq.~\eqref{barotropic_relation}, we assume that the dark matter BEC behaves as a self-interacting polytropic fluid, with pressure depending on density as~\cite{bahcall1980, jimenez2002, weber2009, marsch2011}
\begin{equation} \label{polytropic_eq}
p = U^2\,\rho,
\end{equation}
where $U$ is a constant with dimensions of a velocity, known as the isothermal sound speed~\cite{christodoulou}. Here, we parametrize the relation between the bulk density and pressure as
\begin{equation}
p = U^2\,\rho = \rhO\,U^2\,\Theta(r).
\end{equation}
Defining the galactic rotational period $T = 2\pi/\Omega$, we introduce the parameter
\begin{equation}
\omega = \sqrt{2}\,\Omega\,\tau = \frac{374{\rm~My}}{T},
\end{equation}
whose order of magnitude of the parameter is approximately one. Defining
\begin{equation}
\tau = \left(4\pi G\,\rhO\right)^{-1/2},\quad\hbox{and}\quad q_U = \frac{U\,\tau}{\rO},
\end{equation}
and switching to the new variable $q = r/\rO$, the Lane-Emden Eq.~\eqref{lane_emden} is rewritten as
\begin{equation} \label{lane_emden1}
\Theta + \frac{q_U^2}{q}\,\frac{d}{dq}\,\left(\frac{q}{\Theta}\,\frac{d\Theta}{dq}\right) = \omega^2.
\end{equation}
In the following, we solve Eq.~\eqref{lane_emden1} with the boundary conditions that $\Theta(q)$ and its derivative match the corresponding quantities from the Einasto profile at the solar neighborhood:
\begin{equation} \label{boundary_conditions}
\Theta\left(1\right) = \Theta_E\left(1\right) = 1,\quad\hbox{and}\quad \frac{d\Theta(q)}{dq}\bigg|_{q=1} = \frac{d\Theta_E(q)}{dq}\bigg|_{q=1} = -\frac{2}{q_s^\delta}.
\end{equation}

\subsection{Non-rotating halo $\omega = 0$}

\begin{figure}[t!]
\centering
\includegraphics[height=7cm]{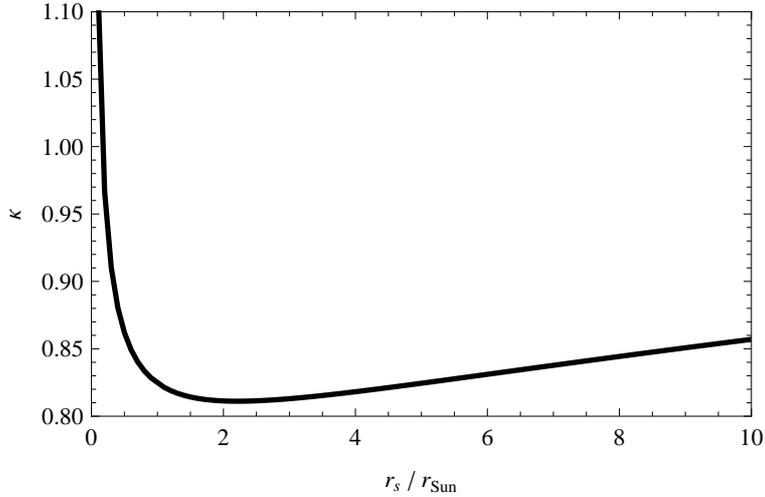}
\caption{The parameter $\kappa$ as a function of $q_s = r_s / \rO$, as given in Eq.~\eqref{relation_kappa_qs}.}
 \label{Fig_kappa}
\end{figure}

When $\omega = 0$, the solution to the differential Eq.~\eqref{lane_emden1} reads
\begin{equation} \label{theta_sol_1p}
\Theta(q) = q^{\kappa-2}\,\left(\frac{q_s^\kappa +1}{q_s^\kappa+q^\kappa}\right)^2, 
\end{equation}
where $q_s$ is related to $q_U$ by 
\begin{equation}
q_s^\kappa = \kappa^2\,q_U^2-1-\kappa\,q_U\,\sqrt{\kappa^2\,q_U^2-2}, \quad\hbox{or}\quad q_U = \frac{q_s^\kappa+1}{\kappa\,\sqrt{2q_s^\kappa}}.
\end{equation}
The expression in Eq.~\eqref{theta_sol_1p} ensures that $\Theta(q) > 0$ for any value of $q$, and satisfies the normalization $\Theta(1) = 1$. The halo profile in Eq.~\eqref{theta_sol_1p} can be rephrased in terms of the function defined in Eq.~\eqref{profile_parametrized} as
\begin{equation} \label{profile_parametrized1}
\Theta(q, \kappa) = \Theta_{\rm CDM}(q, q_s, \kappa,2+\kappa,2-\kappa).
\end{equation}
Notice that, for $\kappa=1$, Eq.~\eqref{theta_sol_1p} reduces to the NFW profile given in Eq.~\eqref{profile_NFW}. We have thus obtained the important result that the NFW profile can be obtained as a solution to the non-rotating Lane-Emden equation. We search for the value of the parameter $\kappa$ which is consistent with the boundary condition for the derivative of the halo profile, as given in Eq.~\eqref{boundary_conditions}. This yields to a transcendental expression for $\kappa$,
\begin{equation} \label{relation_kappa_qs}
\frac{(2-\kappa)(1+q_s^\kappa) + 2 \kappa}{1 + q_s^\kappa} = \frac{2}{q_s^\delta},
\end{equation}
which has to be solved numerically. Setting for example $q_s = (20{\rm kpc})/\rO$, we obtain $\kappa=0.81$, which is the value we adopt later in Sec.~\ref{omega_neq_zero}. In Fig.~\ref{Fig_kappa}, we plot the value of $\kappa$ obtained from Eq.~\eqref{relation_kappa_qs} as a function of $q_s$.

\begin{figure}[t!]
\centering
\includegraphics[height=10cm]{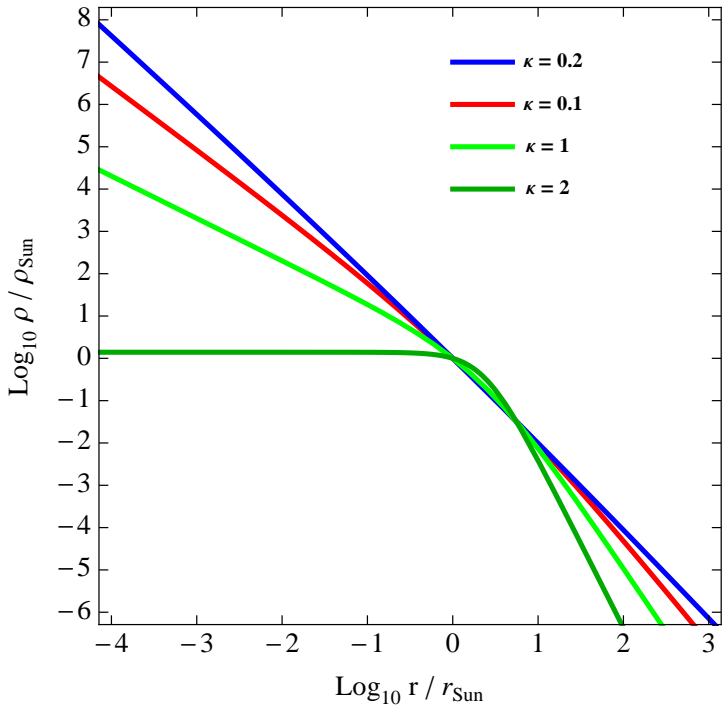}
\caption{The halo density profile $\Theta(q) = \rho/\rhO$ defined in Eq.~\eqref{profile_parametrized1}, as a function of the parameter $q = r/\rO$ with the values $\kappa = 0.2$ (blue solid line), $\kappa = 0.5$ (red), $\kappa = 1$ (green), and $\kappa = 2$ (dark green).}
\label{Fig_Theta}
\end{figure}

Fig.~\ref{Fig_Theta} shows the halo density profile, in units of $\rhO$, as a function of the distance form the galactic center $q = r/\rO$. We set $\kappa = 0.2$ (blue solid line), $\kappa = 0.5$ (red line), $\kappa = 1$ (green line), and $\kappa = 2$ (dark green line). For any value of $\kappa$, all functions drop to zero for large values of $q$. At the halo core, the functions tends to infinity for $0<\kappa<2$, with the steepness of the halo density near the galactic center decreases for increasing values of $\kappa$. In the limit $\kappa = 2$, the halo density profile at the halo core reaches a finite value at the galactic center,
$$\rho(q=0, \kappa=2) = \rhO\,\left(\frac{q_s^2 +1}{q_s^2}\right)^2,$$
similarly to the prediction of the isothermal halo example in Eq.~\eqref{profile_isothermal} where $\rho_{\rm iso}(0) = \rhO$. Despite avoiding the cuspy halo problem, the result for $\kappa = 2$ cannot be reconciled neither with observations nor with galactic simulation.

\begin{figure}[t!]
\centering
\includegraphics[height=8cm]{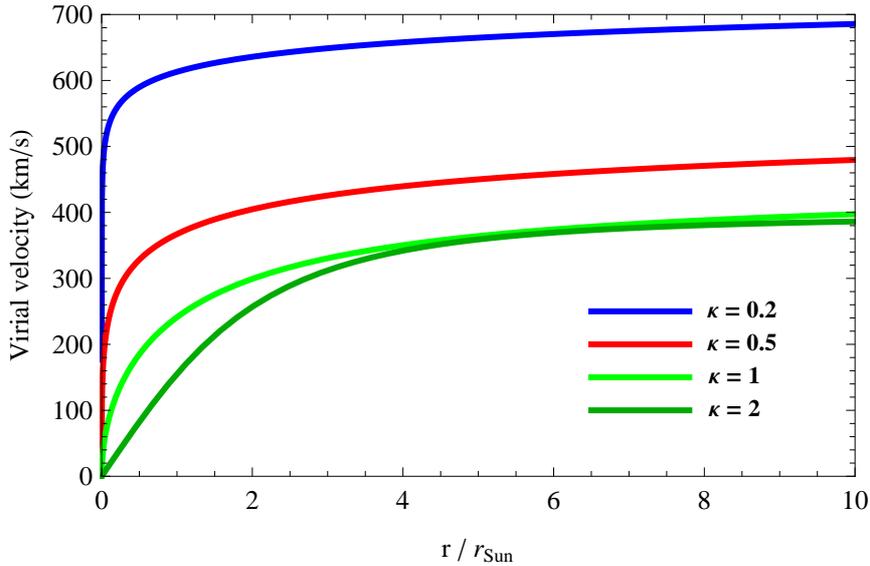}
\caption{The virial velocity (in km/s), given in Eq.~\eqref{virial_vel}, as a function of the parameter $q = r/\rO$. We show results for $\kappa = 0.2$ (blue solid line), $\kappa = 0.5$ (red), $\kappa = 1$ (green), and $\kappa = 2$ (dark green).}
\label{PlotVel}
\end{figure}

The galactic rotational curves follow from the density profile in Eq.~\eqref{theta_sol_1p}. In fact, inserting the density profile into the Poisson Eq.~\eqref{poisson} written in the form
\begin{equation}
\frac{d^2\phi}{dq^2} + \frac{1}{q}\,\frac{d\phi}{dq} = \left(\frac{\rO}{\tau}\right)^2 \Theta(q),
\end{equation}
and integrating twice from $q' = 0$ to $q' = q$, we obtain the expression for the gravitational field
\begin{equation}~\label{potential}
\phi = \vr^2\,\ln\left(\frac{q_s^\kappa}{q^\kappa + q_s^\kappa}\right)^{1/\kappa},
\end{equation}
where
\begin{equation}~\label{virial_constant}
\vr = \frac{\rO}{\tau}\,\frac{1 + q_s^\kappa}{\sqrt{\kappa\,q_s^\kappa}} = 202\,\frac{1 + q_s^\kappa}{\sqrt{\kappa\,q_s^\kappa}}{\rm~km/s}.
\end{equation}
Notice that, for large values of $q \gg q_s$, the potential in Eq.~\eqref{potential} has the analytical expression given in Ref.~\cite{sikivie1998}
\begin{equation}~\label{potential_approx}
\phi \approx \vr^2\,\ln\frac{q_s}{q},\quad \hbox{for $q \gg q_s$}.
\end{equation}
The virial velocity is given by equating $\n \phi$ with the centrifugal force and reads
\begin{equation} \label{virial_vel}
v = \vr\,\sqrt{1-\frac{q_s^\kappa}{q^\kappa + q_s^\kappa}},
\end{equation}
where $\n\phi$ is the gradient of the potential expressed in cylindrical coordinates. We clearly obtain $v \approx \vr$ for $q \gg q_s$. In Fig.~\ref{PlotVel} we plot the virial velocity from Eq.~\eqref{virial_vel} as a function of $q$ for $\kappa = 0.2$ (blue solid line), $\kappa = 0.5$ (red line), $\kappa = 1$ (green line), and $\kappa = 2$ (dark green line). We obtain that the rotational curves rise from the value $v = 0$ at the origin, to reach a flat shape for $r > \rO$. 

\subsection{Rotating halo $\omega \neq 0$} \label{omega_neq_zero}

\begin{figure}[h!]
\centering
\includegraphics[width=15cm]{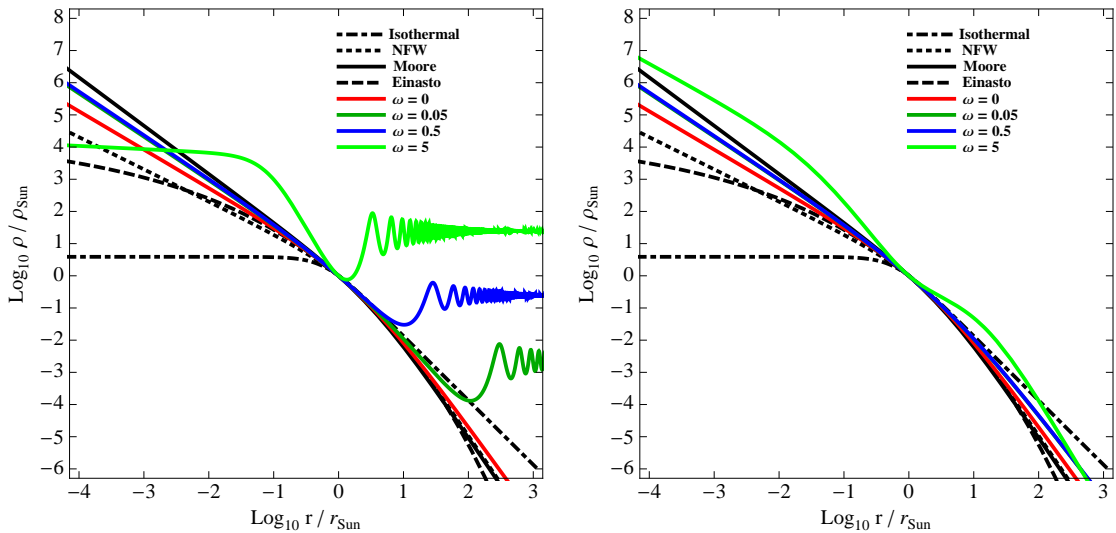}
\caption{(Left) The function $\Theta(q)$ obtained from solving Eq.~\eqref{lane_emden1} with the radius-dependent angular momentum given in Eq.~\eqref{f_omega}, and with the values $\omega = 0$ and $\kappa = 0.81$ (red solid line), $\omega = 0.05$ (dark green), $\omega = 0.5$ (blue), and $\omega = 5$ (light green). For comparison, we have also included the profiles obtained from the isothermal (dot-dashed line), NFW (dotted line), Moore(dashed line), and Einasto profiles (dashed line). (Right) The same figure as shown in the left panel, for an angular momentum depending on radius as in Eq.~\eqref{f_omega}.}
 \label{Fig_Theta1}
\end{figure}

For $\omega \neq 0$, we solve Eq.~\eqref{lane_emden1} numerically with the boundary conditions given in Eq.~\eqref{boundary_conditions}. Fig.~\ref{Fig_Theta1} (left panel) shows the value of $\Theta(q)$ for $\omega = 0$ (solid red), $\omega = 0.05$ (dark green), $\omega = 0.5$ (blue), and $\omega = 5$ (light green). For comparison, we have included the results for the isothermal profile (dot-dashed), the NFW profile (dotted), Moore profile (solid) and Einasto profile (dashed).

The numerical resolution in the left panel of Fig.~\ref{Fig_Theta1} shows an unrealistic behavior for large radii $r \gg \rO$, where the halo density profile exponentially drops to the constant value $\omega^2$, instead of dropping to zero as for the case of the Einasto profile, the NFW profile, and the non-rotating solution. This behavior is due to the fact that, for $r \gg \rO$, the derivative term in Eq.~\eqref{lane_emden1} drops to zero as $1/r^2$, and the equation reduces to $\Theta = \omega^2$. We have further checked that, in this limit, expanding the exact solution $\Theta$ into a perturbation series, the first order perturbation also explains the exponentially damped oscillations observed in the full solution in Fig.~\ref{Fig_Theta1}. To sum up, since $\Theta = \omega^2$ for large values of $r$, we have obtained an unrealistic result because the angular velocity $\OO$ in Eq.~\eqref{lane_emden1} is assumed constant. To circumvent this problem, we modify the Lane-Emden equation by assuming that the angular velocity depends on $q$ as
\begin{equation}
\Omega(q) = f(q)\,\Omega_0, \quad\hbox{and}\quad \omega_0 = \sqrt{2}\,\Omega_0\,\tau,
\end{equation}
where $\Omega_0$ is the value of $\Omega$ when $q = 0$, and $f(q)$ is a generic function of $q$ satisfying~\footnote{A similar parametrization has been used in Ref.~\cite{christodoulou}}
\begin{equation} \label{condition_f}
f(0) = 1\quad\hbox{and}\quad \lim_{q\to+\infty}f(q) = 0.
\end{equation}
The Lane-Emden Eq.~\eqref{lane_emden1} is thus modified as
\begin{equation} \label{lane_emden2}
\Theta + \frac{q_U^2}{q}\,\frac{d}{dq}\,\left(\frac{q}{\Theta}\,\frac{d\Theta}{dq}\right) = f^2(q)\,\omega_0^2.
\end{equation}
The results shown on the right panel of Fig.~\ref{Fig_Theta1} are then obtained for the constant function $f(q) = 1$ . Here, we use the sample function
\begin{equation} \label{f_omega}
f(q) = \frac{1}{1+q^2},
\end{equation}
which satisfies the conditions in Eq.~\eqref{condition_f}, to obtain a solution to Eq.~\eqref{lane_emden2} that fits the numerical results at large radii, as shown in the right panel of Fig.~\ref{Fig_Theta1}.

At the halo core, the numerical resolution of Eq.~\eqref{lane_emden1} grows indefinitely for $q \to 0$, for $\omega < 1$ and $f(q) = 1$, and for any value of $\omega$ and $f(q)$ as in Eq.~\eqref{f_omega}. This behavior is similarly to what obtained for the NFW, Moore, and the halo profile with $\omega = 0$ shown in Fig.~\ref{Fig_Theta}. Within the framework of Eq.~\eqref{lane_emden}, the divergence of $\rho$ at the halo core has been addressed by Christodoulou and Kazanas~\cite{christodoulou}, who suggest to use a composite model in which the hydrostatic solution only applies at large radii, while the halo distribution is truncated to a constant value at small radii. Here, we suggest a different solution which involves the addition of a quantum pressure term, as discussed in Section~\ref{BEC_CDM}.

\section{Equations for the halo profile including quantum pressure corrections} \label{BEC_CDM}

In the literature, the set of Eqs.~\eqref{newton1}-\eqref{poisson} describes classical fluid dynamics. An analogous set of equations might be derived in the context of BEC cold dark matter starting from the Gross-Pitaevskii equation~\cite{chen2005, boehmer2007, kain2010, harko2011a, harko2011b, kain2012, harko2012}. The proof considers the Schroedinger equation for the wave function $\psi(\br,t)$ describing a CDM particle,
\begin{equation} \label{schroedinger}
i\frac{\partial}{\partial t}\,\psi(\br, t) = M_\chi\,\left[-\frac{\n^2}{2M_\chi^2} + \phi(\br) + \phi_{\rm rot}(\br) + \phi_\eta(\br) + \frac{\partial F(\rho)}{\partial\rho}\right]\,\psi(\br, t),
\end{equation}
where $M_\chi$ is the mass of the dark matter particle, $\phi(\br)$ the gravitational potential satisfying the Poisson Eq.~\eqref{poisson}, the potential giving the Coriolis and centrifugal forces is
$$\phi_{\rm rot} = -\frac{1}{2}|\OO|^2\,|\br|^2 + 2\OO\cdot\,\vv\times\br,$$
the potential yielding the viscosities is $\phi_\eta = -\eta\,\br\cdot\n\vv$, and $F(\rho)$ a generic function of the number density, $\rho = |\psi(\br,t)|^2$. From this latter expression, it follows that the wave function can be described as
$$\psi(\br,t) = \sqrt{\rho}\,e^{i\,S(\br,t)},$$
where $S(\br,t)$ is the action of the particle. Once the expression for $\psi(\br,t)$ has been substituted into the steady-state regime of the Schroedinger Eq.~\eqref{schroedinger}, the real component of the expression obtained gives the Navier-Stokes Eq.~\eqref{newton1_steady} with an extra ``quantum pressure'' (QP) term appearing on its right-hand side~\cite{boehmer2007, brook2009, kain2010}. Since $\n S$ is the momentum of the particle, setting $\vv = \n S/M_\chi$ gives
\begin{equation} \label{newton1_qp}
\frac{1}{\rho}\,\n\,p + \n\,\left(\frac{v^2}{2} + \phi\right) + \OO \times \OO \times r + 2\,\OO\times \vv = \eta\,\n^2\,\vv + \frac{1}{2M_\chi^2}\,\n\left(\frac{\n^2\sqrt{\rho}}{\sqrt{\rho}}\right),
\end{equation}
where the pressure is given as a function of $F(\rho)$ as
\begin{equation}
p = \rho\,\frac{\partial F(\rho)}{\partial \rho} - F(\rho).
\end{equation}
The dependence of the QP term on the inverse of the CDM particle mass squared is then predicted by the Schroedinger equation~\cite{chen2005}.

\begin{figure}[t!]
\centering
\includegraphics[height=8cm]{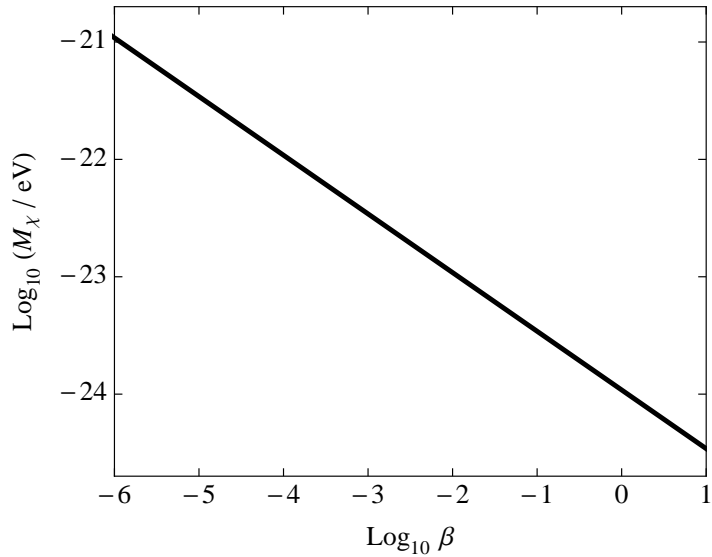}
\caption{The mass of the CDM particle $M_\chi$, as predicted from the model proposed as a function of the parameter $\beta$, see Eq.~\eqref{mass_CDMparticle}.}
 \label{massplot_beta}
\end{figure}

The additional QP term does not modify the barotropic relation $p = p(\rho)$, since it does not appear in the curl of Eq.~\eqref{newton1_qp}. At the same time, the divergence of Eq.~\eqref{newton1_qp} (the Lane-Emden equation) contains an additional term with respect to Eq.~\eqref{bulk_eq},
\begin{equation} \label{div_newton1_qp}
4\pi\,G\,\rho = -\n\left(\frac{1}{\rho}\,\n\,p\right) + 2\Omega^2 + \frac{1}{2M_\chi^2}\,\n^2\left(\frac{\n^2\sqrt{\rho}}{\sqrt{\rho}}\right).
\end{equation}
In cylindrical coordinates, assuming that density depends on the radial coordinate only, and switching to the adimensional variable $q=r/\rO$, Eq.~\eqref{div_newton1_qp} rewrites as
\begin{equation} \label{lane_emden_beta}
\Theta + \frac{q_U^2}{q}\,\frac{d}{dq}\,\left(\frac{q}{\Theta}\,\frac{d\Theta}{dq}\right) = \omega^2 + \frac{\beta}{2q}\,\frac{d}{dq}\,\left\{\frac{q}{\sqrt{\Theta}}\,\frac{d}{dq}\,\left[\frac{1}{q}\,\frac{d}{dq}\,\left(q\,\frac{d\sqrt{\Theta}}{dq}\right)\right]\right\},
\end{equation}
where $\beta$ is an adimensional quantity given by
\begin{equation} \label{beta_to_mass}
\beta = \frac{\tau^2}{\rO^4\,M_\chi^2}.
\end{equation}
Solving Eq.~\eqref{beta_to_mass} for the mass of the CDM particle and reinserting natural units, we find
\begin{equation} \label{mass_CDMparticle}
M_\chi\,c^2 = \frac{\tau\,\hbar\,c^2}{\rO^2\,\sqrt{\beta}} = \frac{1.1}{\sqrt{\beta}}\times 10^{-24}{\rm ~eV},
\end{equation}
a result within the mass range given in Refs.~\cite{marsh2013, hlozek2014, bozek2014, schive2014a, schive2014b, schive2016}. In facts, a CDM particle with a mass in the range $10^{-22}\div10^{24}{\rm~eV}$ is required to solve the cusp problem and to suppress the small-scale spectrum~\cite{hu2000}. Ref.~\cite{lee2009} independently obtain a mass range similar to the one given in this work, in the context of galactic CDM~BEC. We have shown the value of the mass of the CDM particle as a function of $\beta$ in Fig.~\ref{massplot_beta}.

\begin{figure}[t!]
\centering
\includegraphics[height=10cm,width=10cm]{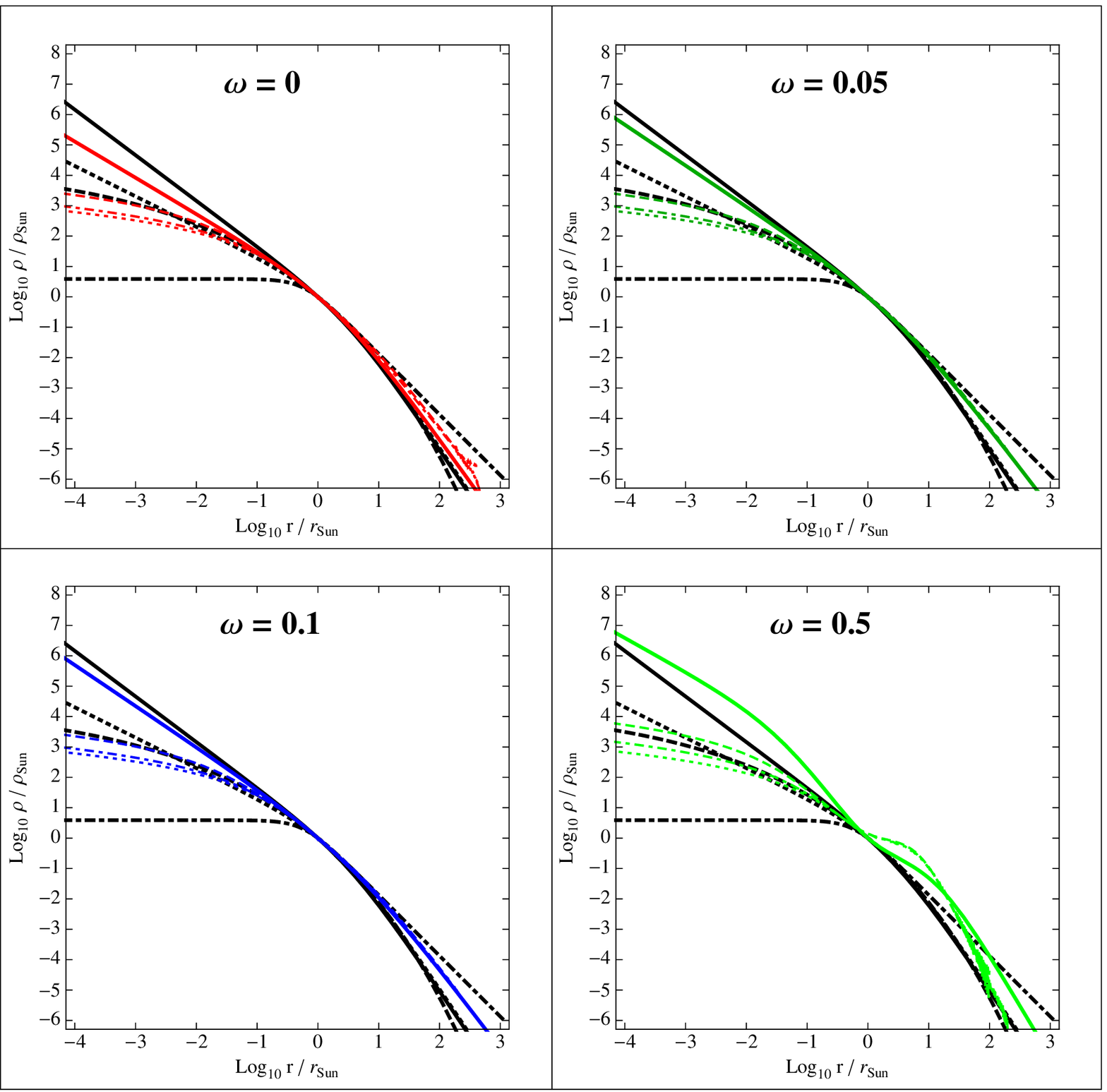}
\caption{The function $\Theta(q)$ solution to Eq.~\eqref{lane_emden_beta}, with the value $\omega = 0$ (top left, red), $\omega = 0.05$ (top right, dark green), $\omega = 0.5$ (bottom left, blue), and $\omega = 5$ (bottom right, light green). For each panel, we have shown the results for $\beta = 0$ (solid line), $\beta = 0.001$ (dashed line), $\beta = 0.01$ (dot-dashed line), and $\beta = 0.1$ (dotted line). Also shown, for comparison, are the NFW profile (solid black line) and the Einasto profile (dashed black line).}
 \label{Fig_Theta_beta}
\end{figure}

Eq.~\eqref{lane_emden_beta} is solved numerically with the boundary conditions in Eq.~\eqref{boundary_conditions}, plus the additional requirement that the second and third derivatives of the numerical resolution match the corresponding quantities in the Einasto profile at the solar neighborhood as well. For the numerical resolution, we have used the radius dependence of the angular momentum given in Eq.~\eqref{f_omega}. The inclusion of QP modifies the solution to the Lane-Emden equation, as we show in Fig.~\ref{Fig_Theta_beta} for the values $\omega = 0$ (top left, red), $\omega = 0.05$ (top right, dark green), $\omega = 0.5$ (bottom left, blue), $\omega = 5$ (bottom right, light green). For each panel, we plot the numerical solutions to Eq.~\eqref{lane_emden_beta} for the values $\beta = 0$ (solid line), $\beta = 0.001$ (dashed line, mass $M_\chi = 3.4 \times 10^{-23}$eV), $\beta = 0.01$ (dot-dashed line, mass $M_\chi = 1.1 \times 10^{-23}$eV), and $\beta = 0.1$ (dotted line, $M_\chi = 3.4 \times 10^{-24}$eV).
\begin{figure}[b!]
\centering
\includegraphics[height=7cm]{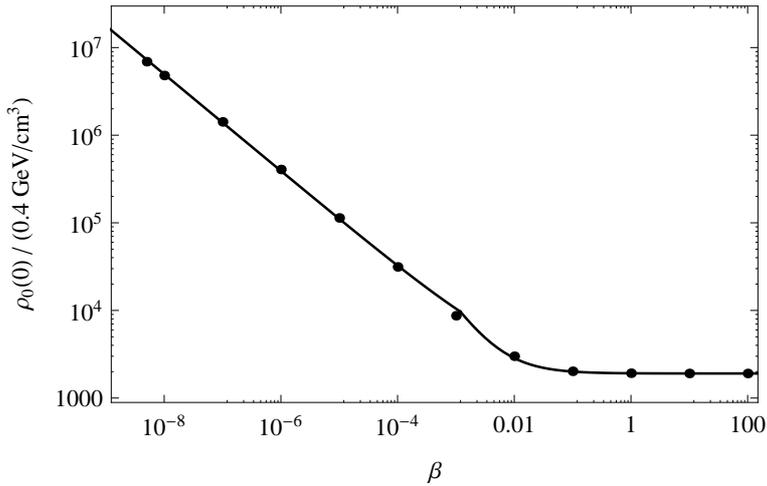}
\caption{Value of $\Theta(q)$ for $q = 0$ for different $\beta$, as obtained from the numerical simulation (dot) and from the fit given in Eq.~\eqref{fit} (solid black line).}
\label{Fig_Theta_beta1}
\end{figure}
For $\beta \neq 0$, the solution to Eq.~\eqref{lane_emden_beta} is finte at the halo core. This result agrees with the assumption made in Ref.~\cite{harko2011b} that the BEC cold dark matter assumption alleviates the cuspy core problem appearing when simulating the evolution of dark matter cores. The stability of such halo, which in principle is not guaranteed~\cite{guzman2013}, will be the subject of a future study.

We fit the results for the value of the halo profile at the halo core from the numerical resolution to Eq.~\ref{lane_emden_beta}, obtaining the result in Eq.~\eqref{fit}. The fit shows a kink at $\bar{q} = 0.0012$, and reaches a constant value for $\beta \to +\infty$, corresponding to a decreasing mass of the DM particle.
\begin{equation} \label{fit}
\Theta(0) = 1907+\begin{cases}
188.55\,\beta^{-0.55}, & q < \bar{q},\\
10.1\,\beta^{-0.99}, & q < \bar{q},
\end{cases}
\end{equation}
In fact, for $\beta \gtrsim 1$, the mass density at the halo core reaches the value
\begin{equation}
\rho(0) = 1907\,\rhO \approx 760{\rm ~GeV/cm^3}.
\end{equation}
We show the fit in Fig.~\eqref{Fig_Theta_beta1}.

In Appendix~\ref{pert_eq}, we derive the expressions for density and pressure perturbations, assuming that they are of the same order of magnitude as the free-streaming velocity $|\vv|$. The solution and discussion to this set of equations will be the subject for further study.

\section{Summary}

Ultra-light scalar dark matter with long-range correlations has been considered as a valid alternative to the ordinary WIMP paradigm. In this paper, we have discussed the halo density profile obtained when treating CDM as a perfect fluid. Our work corroborates the suggestion led by some authors~\cite{harko2011b} that the ``cuspy'' halo core, predicted in ``canonical'' CDM models, is removed when long-range correlations are included. For this, we have first solved the fluid equation for CDM, obtaining that the density profile diverges at the core, see Fig.~\ref{Fig_Theta}. When an extra ``quantum pressure'' term arising from these long-range correlations is included into the fluid equation, the numerically-computed density profile remains finite at the halo core, see Fig.~\ref{Fig_Theta_beta}.

Remarkably, for the parameters $\rhO \approx 0.4{\rm ~GeV/cm^3}$ and $\rO \approx 8.5$~kpc, the model predicts the mass of the CDM particle to be $M_\chi c^2 \approx 1.1 \times 10^{-24}{\rm~eV}/\sqrt{\beta}$, see Eq.~\eqref{mass_CDMparticle}, where $\beta$ parametrizes the correlation length. For values of $\beta \approx O(1)$, Eq.~\eqref{mass_CDMparticle} gives the mass scale of ultra-light scalar CDM, as shown in Fig.~\ref{massplot_beta}.

\begin{acknowledgments}
The author would like to thank the anonymous referee for comments and suggestions, that resulted in a significant improvement over the original manuscript.
\end{acknowledgments}

\section*{Appendix}

\subsection*{Perturbations in the fluid equations} \label{pert_eq}

We linearize the NS Eq.~\eqref{newton1} in the case where the density, pressure, and gravitational potential are perturbed as
\begin{equation} \label{first_order_pert}
\rho = \rho_0 + \rho_1, \quad\quad p = p_0 + p_1, \quad \quad \phi = \phi_0 +\phi_1.
\end{equation}
Substituting this expansion into Eqs.~\eqref{poisson},~\eqref{del_velocity}, and~\eqref{newton1_qp} gives the expression for the zero-th order perturbation as
\begin{eqnarray}
\n\,\phi_0 &=& -\frac{1}{\rho_0}\,\n\,p_0 - \OO \times \OO \times r, \label{NS_bulk2} \\
\frac{d\rho_0}{dt} &=& 0, \label{continuity2} \\
\n^2\phi_0 &=& 4\pi\,G\,\rho_0 \label{poisson2},
\end{eqnarray}
which coincide with the set of Eqs.~\eqref{NS_bulk1}-\eqref{poisson1}. We also obtain the set of equations for the first order perturbations,
\begin{eqnarray}
\frac{1}{\rho_0}\n\,p_1 \!-\! \frac{\rho_1}{\rho_0^2}\n p_0 \!+\! \n\phi_1 \!+\! 2\,\OO\times \vv &=& \eta\,\n^2\,\vv \!+\! \frac{1}{2M_\chi^2}\n\left(\VV_0\,\rho_1 \!+\! \VV_1\,\frac{d\rho_1}{dr} \!+\! \VV_2\,\frac{d^2\rho_1}{dr^2}\right), \label{NS_pert}\\
\n\cdot \vv &=& 0, \label{continuity_pert}\\
\n^2\phi_1 &=& 4\pi\,G\,\rho_1, \label{poisson_pert} \\
\end{eqnarray}
where the coefficients $\VV_0$, $\VV_1$, and $\VV_2$ are obtained from perturbing the term $\n^2(\sqrt{\rho})/\sqrt{\rho}$, and are given by
\begin{align}
\VV_0 &= -\frac{1}{2\rho_0^2}\,\left(\rho_0'' - \frac{(\rho_0')^2}{\rho_0} + 2\frac{\rho_0'}{r}\right), \label{V_0}\\
\VV_1 &= \frac{1}{r\rho_0}\,\left(1 - \frac{r\rho_0'}{2\rho_0}\right), \label{V_1}\\
\VV_2 &= \frac{1}{2\rho_0} \label{V_2}.
\end{align}

The curl of Eq.~\eqref{NS_pert} results in the expression
\begin{equation} \label{NS_curl}
2\n\,\times\,\OO\,\times\,\vv = \eta\,\n^2\,\xx,
\end{equation}
where we introduced the vorticity $\xx = \n\times\vv$. We parametrize the velocity in terms of three new adimensional functions $u$, $v$, $w$, depending on $r$ only, as
\begin{equation}
\vv = \frac{r_s}{\tau}\,\left(u \,\hat{r} + v\,\hat{\phi} + w\,\hat{z}\right).
\end{equation}
Combining the three components of Eq.~\eqref{NS_curl} with the continuity equation gives
\begin{eqnarray}
\frac{d u}{d r} + \frac{u}{r}  &=& 0, \label{NS_curl_u}\\
\nabla^2\,\left(\frac{d v}{d r} + \frac{v}{r}\right) &=& 0, \label{NS_curl_v}\\
\nabla^2\,w -\frac{w}{r^2} &=& 0. \label{NS_curl_w}
\end{eqnarray}
A common solution to the set of Eqs.~\eqref{NS_curl_u}-\eqref{NS_curl_w} that avoids a divergence at infinity is $\vv = \vv_0/r$, for a constant vector $\vv_0$.

We now derive the expression for the divergence of Eq.~\eqref{NS_pert}. Using the barotropic relation in Eq.~\eqref{polytropic_eq}, the relation between pressure and density perturbations is
\begin{equation}
p_1 = U^2\,\rho_1,
\end{equation}
or, writing the series expansion of the function $\Theta = \Theta_0 + \Theta_1$, where $\Theta_0$ is the solution to Eq.~\eqref{lane_emden_beta} and $\Theta_1$ a small perturbation, we obtain
\begin{equation}
\rho_1 = \rhO\,\Theta_1, \quad \hbox{and}\quad p_1 = U^2 \, \rhO\,\Theta_1.
\end{equation}
Using the continuity Eq.~\eqref{continuity_pert}, the divergence of the dissipation term is $\n^2\,\left(\n\cdot\vv\right) = 0$. Once the Poisson Eq.~\eqref{poisson_pert} is taken into account, the divergence of Eq.~\eqref{NS_pert} is a differential equation for $\Theta_1$,
$$\frac{d^2 \Theta_1}{dq^2} + \frac{2}{q}\,\left(1-\frac{q\Theta_0'}{\Theta_0}\right)\Theta_1' + \left[\Theta_0 - \frac{2}{q}\,\frac{\Theta_0'}{\Theta_0} + 2\left(\frac{\Theta_0'}{\Theta_0}\right)^2 - \frac{\Theta_0''}{\Theta_0}\right]\Theta_1 = $$
\begin{equation} \label{NS_div1}
= \frac{\beta}{\lambda^2}\,\frac{1}{2\,q\,\Theta_0^4}\,\left(\Op_0\,\Theta_1 + \Op_1\,\Theta_1' + \Op_2\,\Theta_1'' + \Op_3\,\Theta_1^{(3)}+\Op_4\,\Theta_1^{(4)}\right),
\end{equation}
where the expressions for the coefficients $\Op_i$ are
\begin{align}
\Op'_0 &= 6\,(\Theta_0')^3\left(2\,q\,\Theta_0' -3\Theta_0\right) - 21\,\Theta_0\,(\Theta_0')^2\,\Theta_0'' + 20\Theta_0^2\,\Theta_0'\,\Theta_0'' + \\
& + 4\Theta_0^2\,(\Theta_0'')^2 + 2\Theta_0^2\,\left(3q\,\Theta_0' - 2\Theta_0\right)\,\Theta_0''' - q\,\Theta_0^3\Theta_0'''', \label{O0}\\
\Op'_1 &= -6\Theta_0\,(\Theta_0')^2\,\left(2q\,\Theta_0' - 3\Theta_0\right) + 2\Theta_0^2\,\Theta_0''\,\left(7q\,\Theta_0'-5\Theta_0\right) -3q\,\Theta_0^3\,\Theta_0''',\\
\Op'_2 &= \Theta_0^2\,\left[7q\,(\Theta_0')^2 - 4q\,\Theta_0\,\Theta_0'' - 10\Theta_0\,\Theta_0'\right],\\
\Op'_3 &= -\Theta_0^3\,\left(3q\,\Theta_0' - 4\,\Theta_0\right),\\
\Op'_4 &= q\,\Theta_0^4  \label{O4}.
\end{align}

\end{document}